\documentclass[twocolumn]{aastex701}

\usepackage{threeparttable}
\usepackage{verbatim}  
\usepackage{graphicx}                    
\usepackage{color}                       
\usepackage{breakurl}                         
\usepackage{rotating}
\usepackage{hyperref}
\usepackage{siunitx}

\newcommand\adv{{Adv. in Phys.}}%

\begin{document}

\title{Magnetic Topology and Loop Statistics in Observed Coronal Holes Using Potential Field Modeling}

\author[orcid=0000-0002-2655-2108,sname='SGH']{Stephan~G.~Heinemann}
\affiliation{Institute of Physics, University of Graz, Austria}
\affiliation{Department of Physics, University of Helsinki, Finland}
\email[show]{stephan.heinemann@hmail.at}  

\author[orcid=0000-0003-1175-7124, sname='JP']{Jens~Pomoell} 
\affiliation{Department of Physics, University of Helsinki, Finland}
\email{jens.pomoell@helsinki.fi}

\author[orcid=0000-0003-4867-7558, sname='MT']{Manuela~Temmer} 
\affiliation{Institute of Physics, University of Graz, Austria}
\email{manuela.temmer@uni-graz.at}

\begin{abstract}

Potential Field Source Surface (PFSS) models are widely used to study the solar corona and form the basis for solar wind forecasting, yet often fail to reproduce observed properties of coronal holes. We analyze 702 observed coronal holes between 2010 and 2019 and compute corresponding PFSS magnetic field extrapolations to examine their magnetic topology and loop statistics, comparing them with quiet Sun regions. Our goal is to determine how observed coronal holes are represented in a PFSS model and to identify sources of known discrepancies. We find that low-lying loops covering the weak, balanced background field in coronal holes are statistically smaller and narrower than in quiet Sun regions, with a median height strongly correlated to the coronal holes mean magnetic flux density ($cc_{\mathrm{Pearson}} = 0.81$). This suggests that at low altitudes, the coronal hole magnetic topology is primarily governed by its flux density, unlike in quiet Sun regions. Coronal holes also contain loops extending much higher into the corona than typical quiet Sun loops, although it is unclear if these are truly closed or reflect source surface height limitations. Overall, differences in modeled magnetic structures of coronal holes and quiet Sun regions are evident, even when the PFSS model does not indicate any open fields. These results suggest that observed coronal holes correspond to distinct photospheric magnetic structures, and that discrepancies with PFSS models reflect modeling limitations rather than the absence of coronal holes.

\end{abstract}



\section{Introduction} \label{s:intro}
The solar coronal magnetic field is incredibly complex, shaped by a mix of open and closed field lines that span from very small to global scales. Different features determine the coronal magnetic field structure: most simply categorized into coronal holes, active regions and the quiet Sun. Solar coronal holes are long--living, quasi--stable regions of reduced temperature and density \citep[e.g.,][]{2021heinemann_dem} in the solar corona featuring magnetic fields extending into interplanetary space. Along these open field lines, solar plasma is funneled outward and accelerated to speeds of up to about 800 km/s \citep[see reviews by][and references therein]{cranmer2002,cranmer2009}. This fast-moving plasma forms, what we call the high-speed solar wind streams, which are the main driver of minor to moderate geomagnetic disturbances near Earth \citep{wilcox68,1997farrugia,alves06, 2017vrshnak, 2018richardson}. In contrast, the source of the slow solar wind remains an open question \citep[][]{Viall2020,Temmer2023}. The most frequently discussed mechanisms involve magnetic reconnection, including interchange reconnection in the “S-web” model \citep[separatrix and quasi-separatrix web;][]{Antiochos2011}, or reconnection of closed fields observed in active region cusps \citep[][]{2023Chitta}. {Other potential sources have also been proposed \citep[e.g.,][]{Habbal1997,Fisk1998,Ofman2004,Ko2006,Riley2012}}.\\

A natural conclusion is that understanding the structure of the coronal magnetic field is key in determining the properties of the solar wind throughout the heliosphere. This is the underlying principle of many solar wind models and forms the foundation of space weather forecasting. \\

The solar corona is layered, with dense closed magnetic loops in the low corona that give way to streamer cusps and coronal holes to form open radial magnetic fields in the outer corona. This magnetic topology is responsible how the plasma is confined, heated, and ultimately released as the solar wind \citep[e.g., see][]{2014Reale}. By deriving a representation of the coronal magnetic field, we can model and predict the solar wind and its effects. Although physics-based approaches, such as advanced magnetohydrodynamic (MHD) simulations \citep[see \textit{e.g.,}][]{MAS1996,Lionello1998,Mikic99,Linker1999,Sachdeva2019,Sachdeva2021,Perri2022},  exist{, computationally more efficient models are often preferred for studies of the large scale solar wind and space weather forecasting. However, advanced models, such as MHD simulations, are necessary for investigating small-scale processes and accurately modeling large-scale physical behavior.}
Such models include non-linear force free models \citep[such as][]{2007wiegelmann} and potential field source surface (PFSS) models. The PFSS model \citep[][]{1969altschuler_PFSS,1969Schatten} is the most widely used method to estimate the coronal magnetic field from photospheric magnetic field observations due to its simplicity and computational efficiency. In this approach, the corona is assumed to host a static, current-free magnetic field with a pre-determined source surface beyond which the magnetic field is radial and ignores plasma forces, flows, and partition of energy that full MHD models capture. The PFSS model is an integral part of the Wang–Sheeley–Arge \citep[WSA;][]{2000arge} model and other similar semi-empirical approaches that, in turn, form the backbone of modern space weather forecasting models \citep[e.g.,][]{Odstrvcil1999,Odstrcil2003,Pinto2017,pomoell2018}.\\

Although widely used, PFSS-modeled open fields often do not match observed coronal holes \citep[][]{Asvestari2019, Asvestari2023_ISSI2, Heinemann2023, 2024Heinemann_openflux}. \cite{Asvestari2019} showed that using individually determined source surface heights for each coronal hole, rather than fixed values like 2.5 or 2.6 $R_{\odot}$ \citep[][]{McGregor2011}, improves agreement with observations. However, \cite{Asvestari2023_ISSI2} found that even when testing multiple PFSS variants and a thermodynamic MHD model \citep[][]{Mikic99}, none of the models fully reproduced the observed coronal hole structures. \cite{2024Heinemann_openflux} found that the open flux inferred from observed coronal holes does not agree with the PFSS-modeled open flux, although the authors note that unobserved, non-dark sources of open flux and magnetogram calibration may contribute to this mismatch. These studies suggest a more fundamental problem in potential field modeling, which we address in the following.\\

The magnetic structure of coronal holes is dominated by a weak background field, generally located in the interiors of supergranular cells, with concentrated unipolar flux regions (magnetic elements) appearing along the lanes and nodes of the supergranular or magnetic network \citep[][]{2019hofmeister, Heinemann2018b}. These magnetic elements can either form open funnels that channel solar wind and open flux, or connect with nearby opposite-polarity elements to form closed loops \citep[][]{1973dunn, 2001berger, Heinemann2021_farside}. \cite{2004wiegelmann} demonstrated that coronal holes contain a substantial number of low-lying closed loops, and \cite{Heinemann2021_farside} {suggested} that most of these close near or below  chromospheric heights. This is likely a result of the canopy-like expansion of the open-field funnels, which constrains the vertical development of nearby closed structures \citep[e.g.,][]{Wedemeyer-Boehm2009}. As a result, only larger loops within coronal holes extend to higher altitudes, potentially leaving observable signatures in the lower transition region, such as in He \textsc{ii} 304 \AA\ EUV data \citep[e.g.,][]{2013Jin}.\\

In this study, we use a PFSS model to analyze the magnetic structure of observed coronal holes to better understand their three-dimensional configuration and to investigate potential causes of the mismatch between models and observations. A particular focus is placed on low altitudes that are governed by the rapid expansion of open magnetic fields to fill the corona \citep[][]{cranmer2002}.

\section{Methodology}\label{s:methods}

\begin{figure*}[h]
   \centering
   \includegraphics[width=1\linewidth]{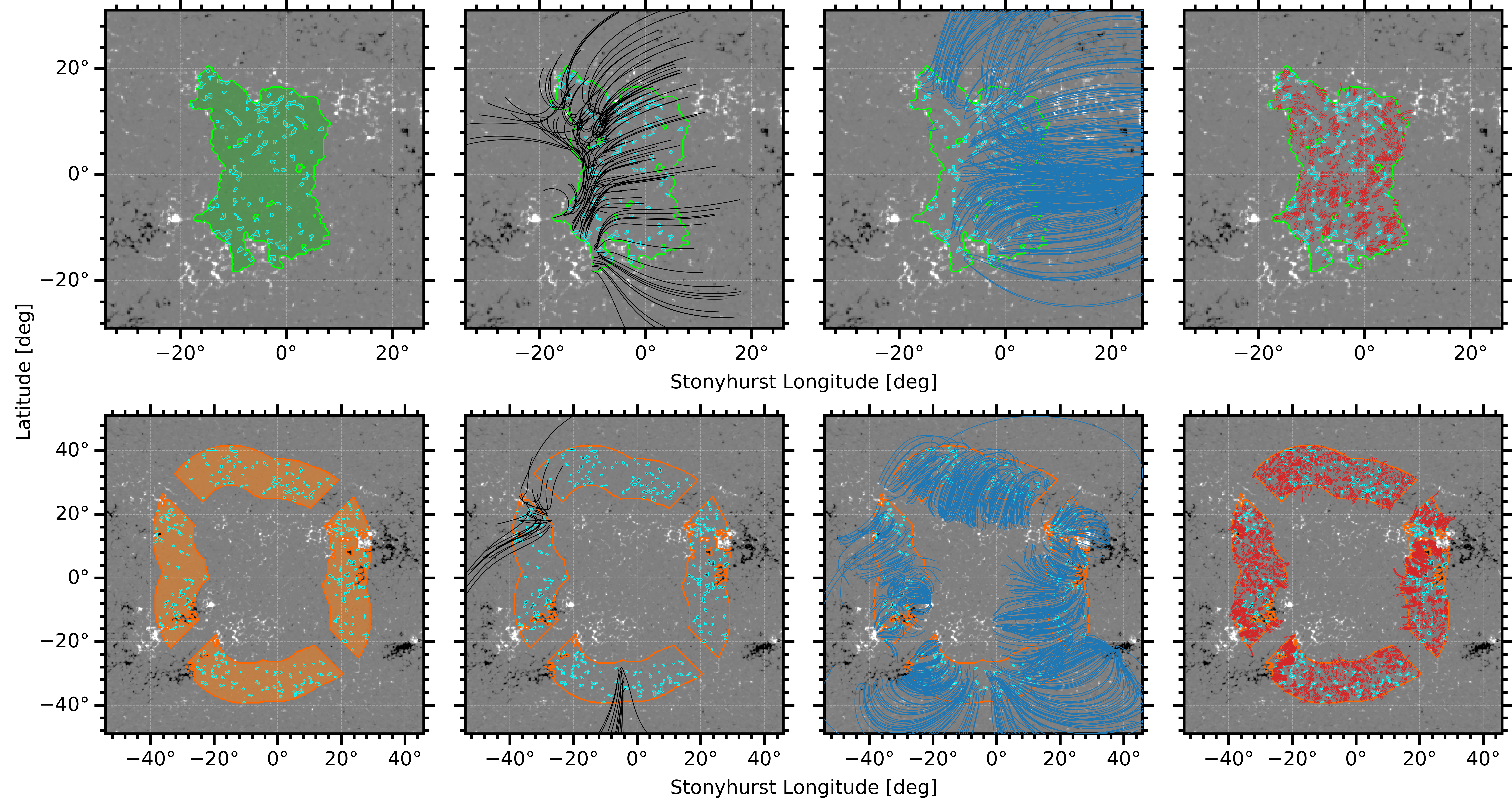}
   \caption{Visual overview of the analysis procedure for a selected coronal hole observed on May 29, 2013. The top row shows the extracted coronal hole, and the bottom row shows the corresponding quiet Sun regions. The second column displays open magnetic field lines (black), the third column shows high loops ($H > 3.48$ Mm; blue), and the last column shows low loops ($H \le 3.48$ Mm; red). Cyan contours denote magnetic elements, highlighting the origins of the different loop structures.}
              \label{fig:overview}%
\end{figure*} 

\subsection{Observed Coronal Holes}
For this study, we utilize the comprehensive, publicly available coronal hole catalog compiled by \cite{2019heinemann_catch}, which includes 707 non-polar coronal holes observed between 2010 and 2019, covering nearly a full solar cycle. These coronal holes were identified using the Collection of Analysis Tools for Coronal Holes (CATCH/pyCATCH) and have been extensively analyzed \citep[\textit{e.g.,}][]{2021heinemann_dem}. The coronal holes were detected in 193~\AA\ filtergrams taken by the Atmospheric Imaging Assembly \citep[AIA;][]{2012lemen_AIA} on-board the Solar Dynamics Observatory \citep[SDO;][]{2012pesnell_SDO} and extracted using a semi-automated threshold method based on the intensity gradient across their boundaries. \cite{2024Reiss} showed that pyCATCH outperforms fully automated models in terms of reliability, producing effectively no false positives, making it well-suited for science-focused studies. The catalog provides both the extracted coronal hole boundaries and associated parameters, such as area and signed and unsigned mean magnetic flux density. For further details, we refer to \cite{2019heinemann_catch}.\\

To construct the PFSS model, an instantaneous full-Sun map of the radial component of the magnetic field at a given base radius is needed. As an approximation to this requirement, we use the \texttt{hmi.mrdailysynframe\_polfil\_720s} daily maps provided by the Joint Science Operations Center\footnote{\href{http://jsoc.stanford.edu/}{http://jsoc.stanford.edu/}}
 (JSOC). This data product combines magnetograms from the Helioseismic and Magnetic Imager \citep[HMI;][]{2012schou_HMI,2016couvidat_HMI} on board SDO to create synoptic charts, where the region within $\pm60^\circ$ longitude as seen from Earth corresponds to the most recent magnetogram. We use the magnetic maps corresponding to the times of the coronal holes in the CATCH catalog (at 12~UT on each given day). {We use this data product, which incorporates an extensive region of the most recent Earth-facing observations, to mitigate the “aging effect” inherent in Carrington map production over a solar rotation. Because some portions of the magnetic maps are considerably older than others, any evolutionary changes in these regions are not captured in the resulting maps. Without concurrent $360^{\circ}$ longitudinal observations of the Sun’s magnetic field, this provides a reasonable approximation, with expected effects primarily on modeled large-scale fields, while small-scale fields remain largely unaffected. Alternatively, flux-transport model maps (e.g., ADAPT: \citealt{Argeetal2010}; FARM: \citealt{Yang2024_FARM}) may be used, though they have their own inherent limitations \citep[][]{Barnes2023}.}

Because HMI data are available only from a later date than AIA, which was used to identify the coronal holes, five coronal holes observed between March and June 2010 lack corresponding synoptic charts and are therefore excluded. This leaves 702 coronal holes for analysis in this study.

\subsection{Potential Field Source Surface Modeling}

We calculate the coronal magnetic field using the PFSS software  \texttt{CIDER}\footnote{\href{https://github.com/jpomoell/cider}{https://github.com/jpomoell/cider}} \citep[e.g. used in][]{2024Heinemann_openflux}. The model employs a finite difference method to numerically solve the Laplace equation for the scalar potential using a staggered grid configuration for the magnetic field. This setup ensures that the magnetic field remains divergence-free and curl-free to within floating-point precision. Additionally, the method accurately reproduces the input magnetic map, enabling fast computation of high-resolution data for the entire coronal hole catalog.

We run the model at a resolution of $0.25^\circ$ per pixel in longitude and latitude, the same resolution to which the magnetic map has been rebinned, and use $100$ radial grid cells with a stretched radial geometry such that the smallest cell, located at lower radial boundary, has an extent of $\Delta r = 0.028$ Mm and the largest $56.11$ Mm. This radially stretched grid resolves loops below chromospheric heights ($<3.5$~Mm; or $<1.005\,R_\odot$) without requiring an excessive number of radial cells. As such we are able to resolve even low-lying loops that close below the point where the open fields or large loops start to expand into the corona. This is thought to happen around the height of the chromosphere ($<3.5$ Mm, or $1.005$~$R_\odot$) \citep[][]{Wedemeyer-Boehm2009,cranmer2009}. For the source surface, we adopt the standard value of 2.5 $R_\odot$ \citep[][]{1969altschuler_PFSS}, commonly used in operational models. While the choice of source surface height is known to affect the representation of open field lines and the global magnetic structure {\citep[e.g.,][]{1983Hoeksema,2014Arden,Asvestari2019,Wagner2022,2024Benavitz}}, we consider this standard value sufficient for the purposes of this study. The primary focus of this study is on the low-lying magnetic field, which is largely unaffected by the choice of source surface height. Appendix Figure \ref{fig:Rss_example} illustrates this by showing the loop statistics of a representative coronal hole for several source surface heights; the distribution of low-lying loops remains essentially unchanged. This behavior is typical of the full dataset.\\

\subsection{Coronal Hole -- PFSS Comparison}
To determine how the observed EUV coronal holes align with the open field regions as determined by the PFSS model, we first trace the magnetic field at a $0.5^\circ$ resolution (in contrast to the $0.25^\circ$ for the loop statistics)  outwards starting at a height of $r = R_1 = 1.01$~$R_{\odot}$ (i.e., from the pixel center of every second pixel). In addition, a second set of field lines are traced inwards starting from the source surface. Pixels are designated as open if they are either start pixels whose field lines reach the source surface, or pixels reached when tracing field lines inwards from the source surface. We then compute the Jaccard Index as
\( J(A, B) = \frac{|A \cap B|}{|A \cup B|} \)
in a region of $\pm 10^\circ$ around the observed coronal hole, but constrained in latitude ($\lambda$) for $|\lambda|<60^\circ$ to avoid polar regions. \\

\subsection{Compute Coronal Hole Loop Statistics}
To analyze the 3D structure and loop statistics of coronal holes, we trace magnetic field lines only from the centers of pixels at $r = R_\odot$ that fall within the observed coronal hole boundaries, rather than also tracing from the source surface inward. For the obtained field lines that do not reach the source surface height (i.e., loops), we calculate their maximum height ($H$, in $R_{\odot}$), loop length ($L$, in $R_{\odot}$), angular separation of the footpoints of the field line ($d$, in degrees), aspect ratio ($a=\frac{H}{d}$, dimensionless), relative length ($L_{\rm rel}=\frac{L}{d}$, dimensionless) and semicircle fit ($f_{\rm circ} = \frac{L}{\pi d/2}$, dimensionless). $H$, $L$ and $d$ are also given in Mm. As reference, we compute the same quantities at the same resolution also for quiet Sun regions, as follows. For each date with a coronal hole, we define four quiet Sun regions located between $35$ and $50$ pixel to the east, west, north, and south of the EUV coronal hole. We exclude all regions that contain at least $30$ pixels (at $0.25^\circ$ resolution) with a magnetic field strength $|B| > 25,\mathrm{G}$ in order to mask out active regions. This results in 2808 quiet Sun regions that were analyzed. An example of the analysis procedure is shown in Figure~\ref{fig:overview}\footnote{*Due to the close resembles to the island, this coronal hole was dubbed "Sardinia" by J. Pomoell.}{, with three additional examples shown in Figures~\ref{fig:overview_app1}, \ref{fig:overview_app2}, and \ref{fig:overview_app3} in the Appendix.}

\section{Results}\label{s:res}

\subsection{Individual Loop Statistics}
To better investigate the magnetic structure, we separate the loops into two domains: low loops with a maximum height of $1.005$~$R_\odot$ ($H\le 3.48$~Mm) and high loops ($H > 3.48$~Mm). In addition, there are the open fields that extend up to the source surface and do not close in the photosphere. Here we note that the open fields{, and to some extent,} the high loops are influenced by the choice of the source surface height but this effect can be shown to be minimal for low loops (see Appendix Fig.~\ref{fig:Rss_example}).\\

\begin{figure*}[h]
   \centering
   \includegraphics[width=1\linewidth]{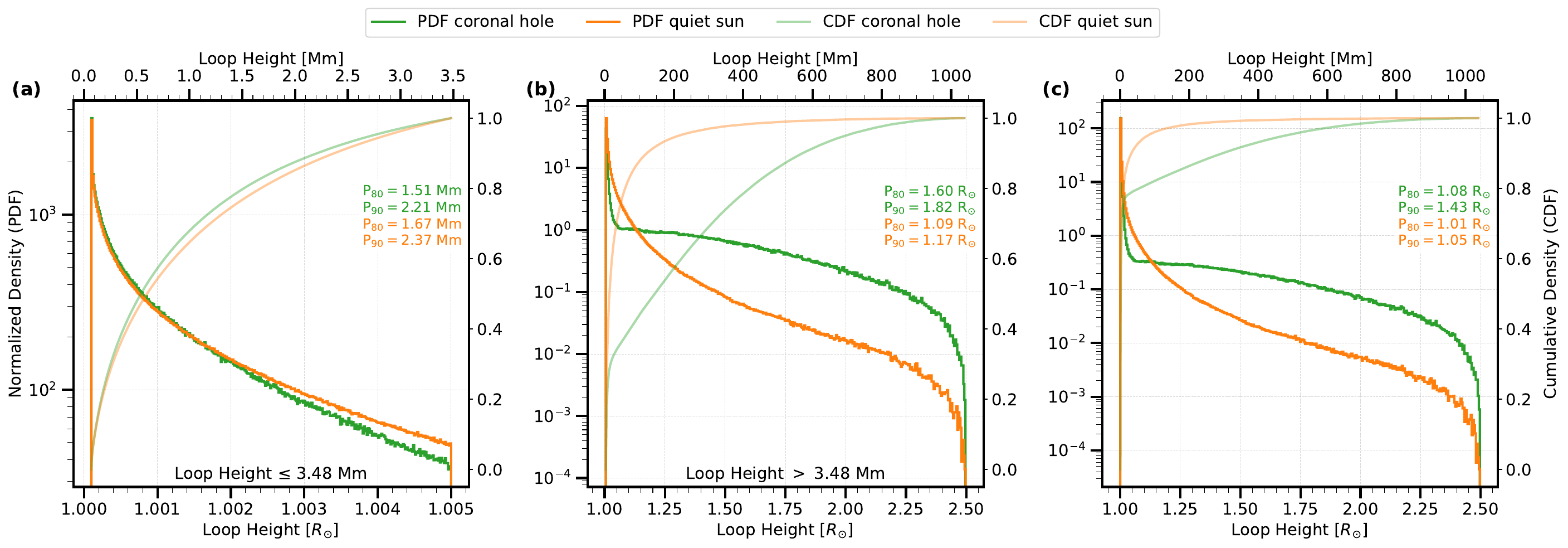}
   \caption{Normalized distribution of the maximum loop height for low (a), high (b) and all loops (c). This is shown for coronal hole loops (green) and quiet Sun loops (orange) with the solid line showing the normalized density (PDF) and the shaded lines showing the cumulative density (CDF). The 80\textsuperscript{th} (P$_{80}$) and 90\textsuperscript{th} (P$_{90}$) percentiles are marked. }
              \label{fig:hist_maxheight}%
\end{figure*}

Figure~\ref{fig:hist_maxheight} shows the distribution of loop heights ($H$) for all coronal holes and quiet Sun loops. For the low loops, we find that quiet Sun regions ({80th percentile, }P$_{80}^{QS}=1.67$~Mm) statistically have more loops closing at higher heights than coronal holes (P$_{80}^{CH}=1.51$~Mm). The distribution of high loops on the other hand shows that coronal holes have significantly more higher loops  (P$_{80}^{CH}=1.60$~$R_{\odot}$) than quiet Sun regions (P$_{80}^{QS}=1.09$~$R_{\odot}$). Considering all loops (excluding open field lines), a significant proportion closes at low heights: about $80\%$ of loops in quiet Sun regions close below coronal heights at around $1.01$~$R_\odot$, whereas in coronal holes, $80\%$ close below $1.08$~$R_\odot$. 

For the distributions of footpoint separation distance ($d$; Fig.~\ref{fig:hist_footpointsep}) and loop length ($L$; Fig.~\ref{fig:hist_looplength}), we find that, similar to the loop heights, low loops in coronal holes tend to be shorter and have more closely spaced footpoints than those in quiet Sun regions. The high-loop distribution in coronal holes exhibits two distinct regimes: loops with $d < 10^{\circ}$ occur with an exponentially decreasing frequency as a function of $d$, whereas loops with $d > 10^{\circ}$ show an approximately constant occurrence rate up to $d = 60^{\circ}$. In contrast, the occurrence of high loops in quiet Sun regions decreases near exponentially with increasing footpoint separation. A very similar behavior is observed in the distribution of high-loop lengths ($L$). In quiet Sun regions, $80\%$ of the loops are shorter than 0.38~$R_{\odot}$, whereas coronal holes contain a larger fraction of much longer loops, with the 80\textsuperscript{th} percentile reaching $P_{80}^{\rm CH} = 1.97$~$R_{\odot}$. The distributions of loop length and footpoint separation distance are shown in Figures~\ref{fig:hist_footpointsep} and \ref{fig:hist_looplength} in the Appendix.\\

\begin{figure}[h]
   \centering
   \includegraphics[width=1\linewidth]{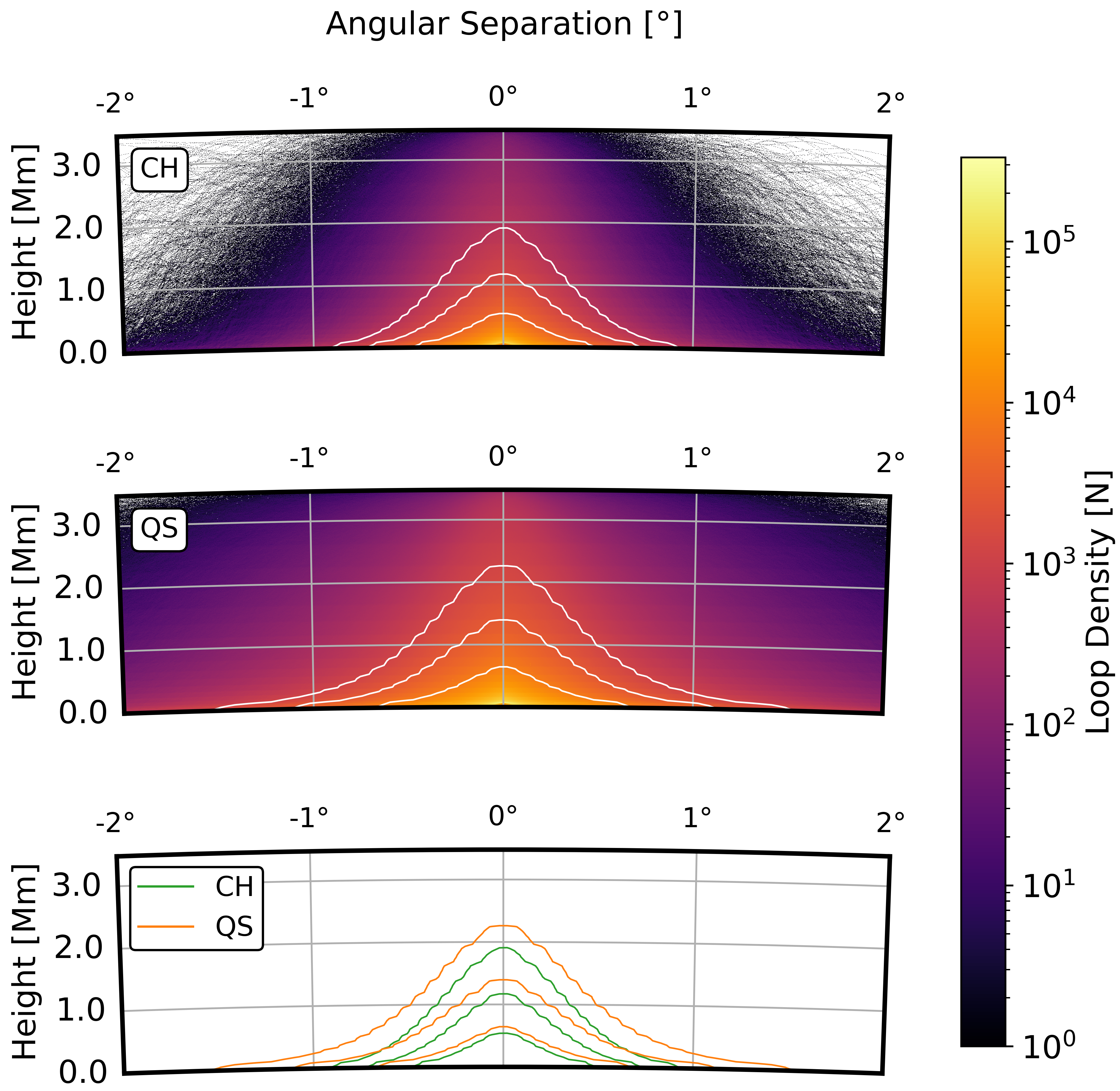}
   \caption{2D distribution of low-lying loops projected onto the baseline connecting the loop footpoints in polar coordinates. The top panel shows the coronal hole loops, the middle panel the quiet Sun loops and the bottom panel the contour lines from both upper panels. The contour lines correspond to the $60$\textsuperscript{th}, $80$\textsuperscript{th}, $90$\textsuperscript{th} percentiles.}
              \label{fig:lowloops}%
\end{figure}

\begin{figure}[h]
   \centering
   \includegraphics[width=1\linewidth]{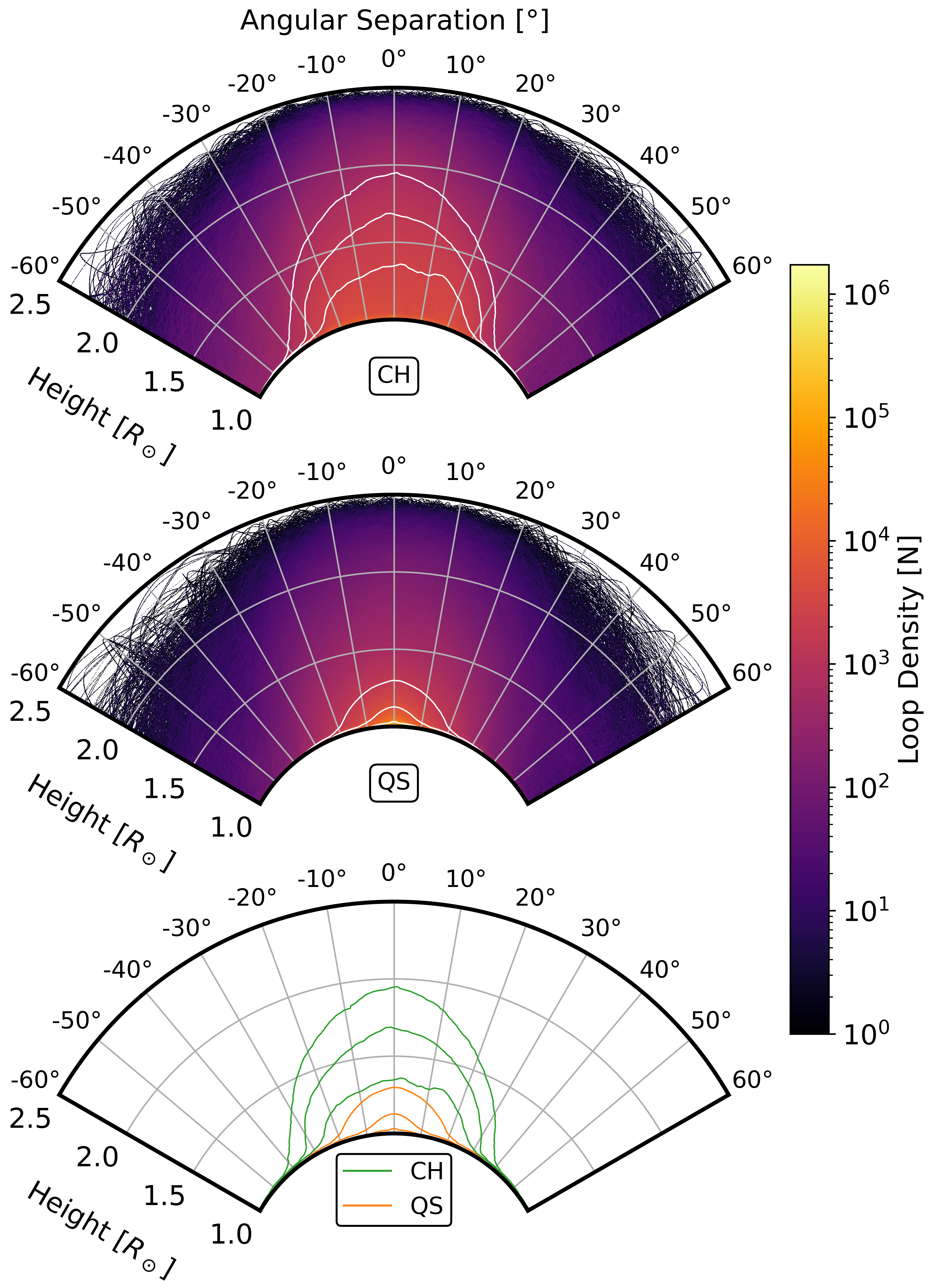}
   \caption{Same as Figure~\ref{fig:lowloops} but for high loops instead.}
              \label{fig:highloops}%
\end{figure}

Furthermore, we visualize these distributions as two-dimensional histograms of all loops in Figure~\ref{fig:lowloops} for the low loops and Figure~\ref{fig:highloops} for the high loops. Each histogram is generated by projecting the individual field lines onto a two-dimensional plane defined by the footpoint baseline and height. Contours corresponding to the 60\textsuperscript{th}, 80\textsuperscript{th}, and 90\textsuperscript{th} percentiles are then computed. For comparison, the bottom panels show the contours from both the coronal hole (top) and quiet Sun (middle) distributions. For the low loops, it is evident that higher and significantly wider loops (i.e., with larger $d$) are more prevalent in the quiet Sun, whereas for the high loops, both higher and wider, as well as longer, structures are more common in coronal holes.

We quantify these results statistically in Table~\ref{tab:combined_stats}. From the $\sim2\times10^{7}$ computed field lines, of which $12.5\%$ were rooted in observed coronal holes and $87.5\%$ in the defined quiet Sun areas, we compute the median as well as the 20\textsuperscript{th} and 80\textsuperscript{th} percentiles of the loop statistics described in Section~\ref{s:methods}. In coronal holes, we find that $61\%$ of the field lines fall into the low loop category, $28\%$ correspond to high loops, and about $11\%$ are open field lines. In quiet Sun areas, $67\%$ of the field lines are classified as low loops and $32\%$ as high loops, with only a minor fraction ($1\%$) being designated as open.

From this, we can quantify the statistical differences between loops modeled in coronal hole and quiet Sun areas. For low loops, we find that those in coronal holes are lower (by $\approx10\%$), shorter (by $\approx15\%$), and narrower (by $\approx15\%$) than their counterparts in quiet Sun regions. However, we do not find significant differences in their shapes, as all areas exhibit similarly wide and flat loops ($a$, $L_{\mathrm{rel}}$, and $f_{\mathrm{semi}}$ are very similar).

In contrast, for high loops we find that coronal holes contain a much larger fraction of loops extending far into the corona. The median height ($H$) is about an order of magnitude larger in coronal holes ($145$~Mm) than in quiet Sun areas ($14$~Mm), with corresponding differences in $d$ (coronal holes: $324$~Mm; quiet Sun: $66$~Mm) and $L$ (coronal holes: $555$~Mm; quiet Sun: $81$~Mm). We also observe pronounced differences in loop shape: the median aspect ratio ($a$) is $\sim60\%$ larger in coronal holes, meaning that loops there are about $60\%$ higher for the same footpoint separation compared to those in quiet Sun regions. Their median shape is close to semicircular ($f_{\mathrm{semi}}\approx1$ and $L_{\mathrm{rel}}\approx\pi/2$), whereas quiet Sun loops are considerably flatter ($f_{\mathrm{semi}}=0.82$ and $L_{\mathrm{rel}}=1.29$), though still distinct from the low-loop population.

\begin{table*}[h!]
\centering
\begin{threeparttable}
\renewcommand{\arraystretch}{1.5}
\begin{tabular}{l|cc|cc|cc}
&\multicolumn{2}{c}{Loop Height $\le$ $3.48$ Mm} & \multicolumn{2}{c}{Loop Height $>$ $3.48$ Mm} & \multicolumn{2}{c}{All Loops} \\
Metric & CH & QS & CH & QS & CH & QS \\
\hline
$H$ [Mm] & $0.54_{0.16}^{1.51}$ & $0.60_{0.17}^{1.67}$ & $145.01_{6.92}^{419.05}$ & $13.72_{5.36}^{59.21}$ & $1.20_{0.24}^{55.57}$ & $1.35_{0.26}^{8.85}$ \\
$d$ [Mm] & $5.25_{2.80}^{9.64}$ & $6.14_{3.09}^{12.06}$ & $324.11_{26.38}^{605.20}$ & $65.55_{23.70}^{183.24}$ & $8.30_{3.46}^{179.85}$ & $10.12_{3.92}^{45.21}$ \\
$L$ [Mm] & $5.61_{2.88}^{10.83}$ & $6.60_{3.18}^{13.30}$ & $555.73_{33.78}^{1376.70}$ & $80.80_{30.17}^{262.60}$ & $9.28_{3.58}^{253.67}$ & $11.36_{4.08}^{54.47}$ \\
$a$\textsuperscript{\textup{i} } & $0.10_{0.05}^{0.18}$ & $0.09_{0.05}^{0.17}$ & $0.44_{0.27}^{0.72}$ & $0.27_{0.17}^{0.41}$ & $0.15_{0.07}^{0.37}$ & $0.13_{0.06}^{0.26}$ \\
$L_{rel}$\textsuperscript{\textup{ii} } & $1.04_{1.01}^{1.12}$ & $1.04_{1.01}^{1.11}$ & $1.62_{1.28}^{2.28}$ & $1.29_{1.15}^{1.54}$ & $1.09_{1.02}^{1.46}$ & $1.08_{1.02}^{1.26}$ \\
$f_{semi}$\textsuperscript{\textup{iii} } & $0.66_{0.65}^{0.71}$ & $0.66_{0.64}^{0.71}$ & $1.03_{0.81}^{1.45}$ & $0.82_{0.73}^{0.98}$ & $0.69_{0.65}^{0.93}$ & $0.69_{0.65}^{0.80}$ \\
$R_i$ [$\%$] & $60.84$ & $67.04$ & $28.47$ & $31.81$ & $89.31$ & $98.85$ \\
$R_{loops}$\textsuperscript{\textup{iv} } [$\%$] & $7.55$ & $58.72$ & $3.53$ & $27.87$ & $11.08$ & $86.59$ \\
\hline
\end{tabular}
\caption{Field line statistics: each kind (low/high/all) with CH and QS. Median with 20th and 80th percentiles.}
\label{tab:combined_stats}
\begin{tablenotes}
\tiny
\item[\textup{i}] Aspect ratio $a = H / d$; $a\ll 1$ → flat loops, $a > 1$ → tall loops.
\item[\textup{ii}] Relative length $L_{rel} = L / d$;  $L_{rel} \approx 1$ → almost straight, $L_{rel}>1$ → curved/elongated; (semicircle $\approx\pi/2$).
\item[\textup{iii}] Semicircle fit $f_{semi} = L / (\pi \, d/2)$;  $f_{semi}\approx1$ → semicircular, $f_{semi}<1$ → flatter, $f_{semi}>1$ → stretched.
\item[\textup{iv}] Percentage of loops among the total calculated field lines.
\end{tablenotes}
\end{threeparttable}
\end{table*}

\subsection{Region Based Loop Statistics} \label{subs:region}
We then computed the loop statistics for each region (coronal hole, quiet Sun) separately and related these to the properties of each given region. These are area ($A$), absolute value of the signed mean magnetic flux density ($|B_{\mathrm{s}}| =|\frac{\Phi_s}{A}|$) and loop ratio (defined as $R_{i}=\frac{N_{i}}{N_\mathrm{low}+N_\mathrm{high}+N_\mathrm{open}}$, where $N_{i}$ is either $N_\mathrm{low}$, $N_\mathrm{high}$ or $N_\mathrm{open}$). For coronal holes, $A$ and $|$B$_{\mathrm{s}}|$ were taken from the CATCH catalog\footnote{\href{https://vizier.cds.unistra.fr/viz-bin/VizieR?-source=J/other/SoPh/294.144}{https://vizier.cds.unistra.fr/viz-bin/VizieR?-source=J/other/SoPh/294.144}} and for the quiet Sun areas computed following the description in \cite{Heinemann2018b,Heinemann2018a,2019heinemann_catch,Hofmeister2019}. 

\begin{figure*}[h]
   \centering
   \includegraphics[width=1\linewidth]{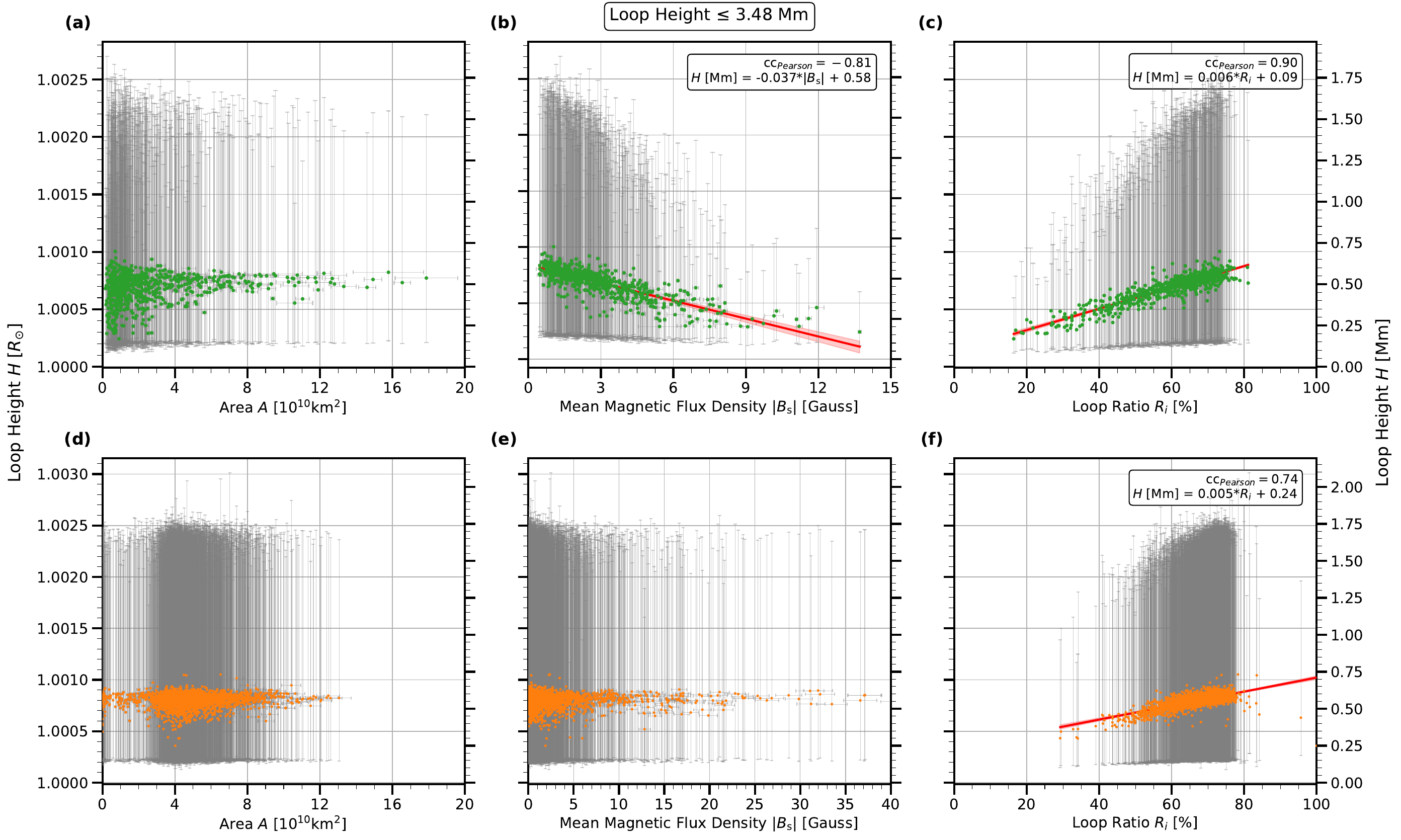}
   \caption{Median height of low loops ($H \le 3.48$~Mm) as a function of the respective coronal hole (a–c; green) and quiet Sun (d–f; orange) parameters: area ($A$), absolute value of the signed mean magnetic flux density ($|$B$_{\mathrm{s}} |$), and loop fraction ($R_i$). Grey bars indicate the 20\textsuperscript{th} and 80\textsuperscript{th} percentiles of each region. Linear fits are shown for cases with Pearson correlation coefficients exceeding $0.5$.}
              \label{fig:metric_maxheight_low}%
\end{figure*}

\begin{figure*}[h]
   \centering
   \includegraphics[width=1\linewidth]{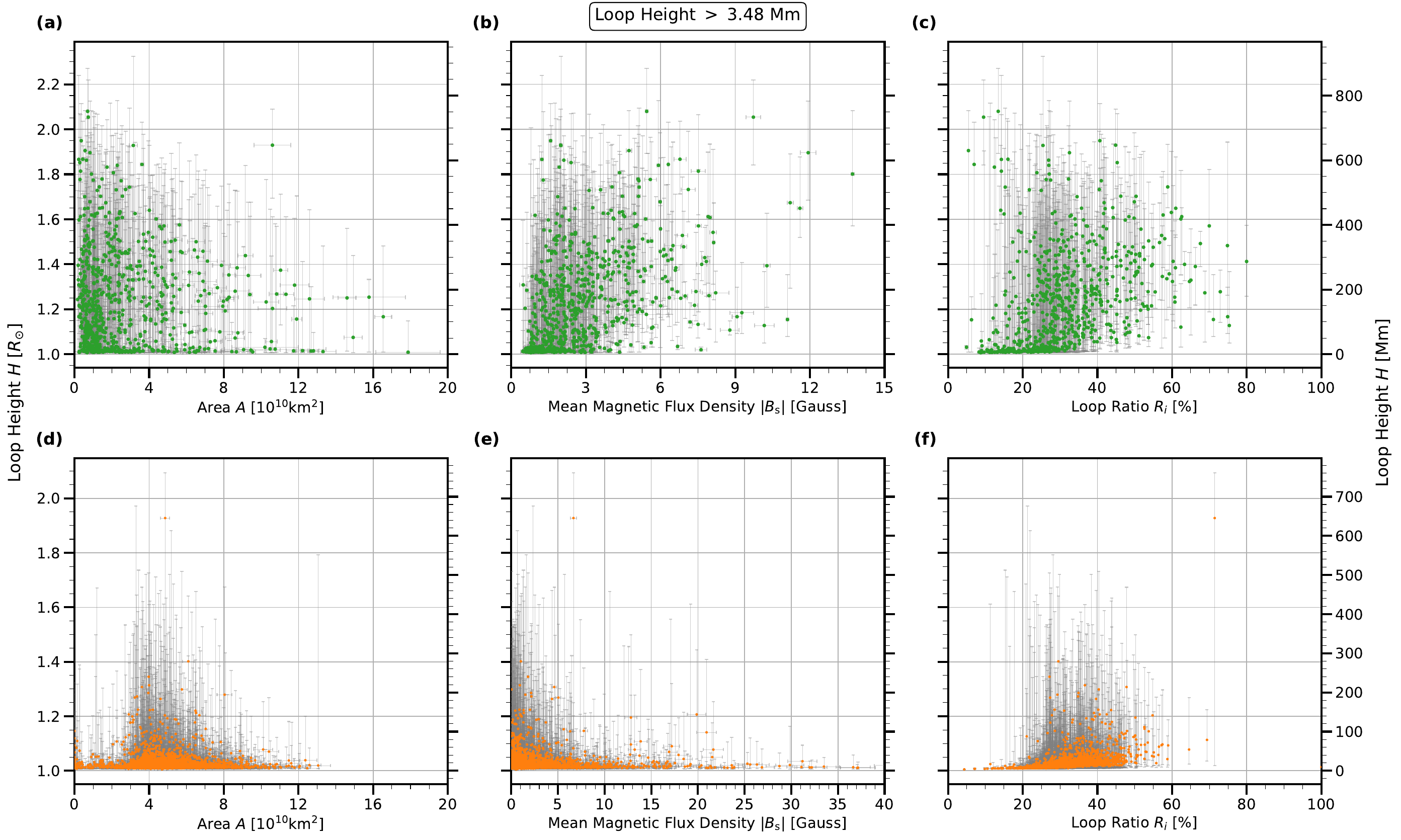}
   \caption{Same as Figure~\ref{fig:metric_maxheight_low} but for high loops ($H > 3.48$~Mm). }
              \label{fig:metric_maxheight_high}%
\end{figure*}

The distributions of $H$ are shown as scatter plots in Figures~\ref{fig:metric_maxheight_low} and~\ref{fig:metric_maxheight_high} for low and high loops, respectively. The corresponding quantities $d$ and $L$ are presented in Figures~\ref{fig:metric_fp_distance_low},~\ref{fig:metric_fp_distance_high},~\ref{fig:metric_fl_length_low}, and~\ref{fig:metric_fl_length_high} in the Appendix.

For low loops, we find that in coronal holes the median loop height is strongly anti-correlated with the mean magnetic flux density ($| B_{s} |$), with a Pearson correlation coefficient of $cc=-0.81$ (Fig.~\ref{fig:metric_maxheight_low}b). In contrast, no such correlation is found in quiet Sun regions (Fig.~\ref{fig:metric_maxheight_low}e). The median $H$ also appears to depend on the fraction of low loops present: a higher fraction of low loops corresponds to a larger median $H$ in both coronal holes ($cc=0.90$, Fig.~\ref{fig:metric_maxheight_low}c) and quiet Sun areas ($cc=0.74$, Fig.~\ref{fig:metric_maxheight_low}f). We find no relationship between $H$ and the corresponding region size $A$ for low loops.

For high loops (Fig.~\ref{fig:metric_maxheight_high}), we do not find any significant correlations between $H$ and the associated region parameters.

Additionally, in quiet Sun areas, the footpoint separation distance ($d$) is anti-correlated with the fraction of low loops ($cc=-0.64$, Fig.~\ref{fig:metric_fp_distance_low}f) and correlated with the fraction of high loops ($cc=0.60$, Fig.~\ref{fig:metric_fp_distance_high}f), but shows no correlation with either $A$ or $|B_{s}|$. The same behavior is observed for $L$, with $cc=-0.63$ (Fig.~\ref{fig:metric_fl_length_low}f) and $cc=0.55$ (Fig.~\ref{fig:metric_fl_length_high}f) for low and high loops, respectively. For coronal holes, we do not find any significant correlations between $d$ or $L$ and the corresponding region parameters.\\

As our dataset covers nearly a full solar cycle, we analyzed the loop characteristics as a function of time (Figs.~\ref{fig:dt_low} and~\ref{fig:dt_high}). We find that in coronal holes, the loop statistics show no clear relation to the solar cycle and remain largely stable over time, although considerable variability is observed in the properties of high loops. In contrast, quiet Sun regions exhibit a stronger temporal dependence, particularly for low loops. Specifically, the median values of $d$ and $L$ for low loops in quiet Sun regions show pronounced variations during 2013–2016, while high loops in these regions remain relatively uniform throughout the cycle.

Furthermore, we computed overlap metrics to assess how well the PFSS model with the chosen source surface height reproduces the observed coronal holes. We find that the agreement between the modeled and observed open field regions is generally poor, with an average Jaccard Index of $0.13 \pm 0.14$ ($0.18 \pm 0.13$ when considering only events with $J>0$) and a maximum of $0.67$. For 185 coronal holes ($26.4\%$ of the events), the model did not produce any open field regions anywhere within {the observed coronal hole areas}.

\section{Discussion}\label{s:disc}

In this study, we investigated the magnetic field structure of coronal hole loop systems under the assumption of a potential field in comparison to quiet Sun regions.

We find that the PFSS-modeled open magnetic field shows poor agreement with the observed coronal holes {(e.g., see Fig.~\ref{fig:overview_app1})}. This discrepancy is partly attributable to the standard choice of the source surface height at $2.5$~$R_{\odot}$. Adjusting the source surface height manually can improve the correspondence with observations \citep[e.g.,][]{Asvestari2019, Asvestari2023_ISSI2, Linker2021, Heinemann2023}; however, such tuning does not guarantee that other regions in the corona are represented correctly. To address the limitation of using a single global source surface height, several studies have proposed non-spherical source {surfaces \citep[e.g.,][]{Schulz1978,1979Sakurai,1981Sakurai,1982Levine_Sourcesurface, 2015Cohen, 2020Panasenco}. For example,} \citet{2020Kruse} implemented an elliptic expansion of the source surface but did not evaluate its performance against coronal observations or solar wind measurements to demonstrate any improvement.

For the present study, the precise agreement between the PFSS modeled open field and coronal hole observations is not of primary importance. Nevertheless, it implies that it remains unclear, and currently unreliable, to determine which of the traced high loops in the coronal hole are truly open magnetic structures on the Sun, that are merely modeled as closed with the chosen source surface height. 

Supporting this interpretation, we find that the median height of high loops in the model with a footpoint within coronal holes is significantly greater than in quiet Sun regions {(Fig.~\ref{fig:hist_maxheight}b)}. These loops often connect to magnetic concentrations far from the coronal hole boundary, whereas high loops in quiet Sun areas typically connect to nearby opposite-polarity regions. From these results, we conclude that while the {modeled} magnetic structure in the quiet Sun appears relatively uniform {(note that for PFSS models: $\nabla \times {B} = 0$)}, this is not the case for coronal holes (and likely not for active regions either). This finding supports the idea of a non-uniform, spatially varying source surface height that depends on the magnetic source region, that can potentially improve the realism of solar and heliospheric magnetic field models and space-weather forecasting capabilities.\\

When investigating the magnetic field structure at low altitudes, up to approximately the chromosphere ($\leq 1.005$~$R_{\odot}$; $\leq 3.48$~Mm), we find that the median loop footpoint separation distance is about 5–6~Mm in both coronal holes and quiet Sun regions {(Fig.~\ref{fig:hist_footpointsep}a)}. This scale corresponds to the size of the magnetic internetwork observed by Hinode \citep[][]{2008Lites}. The slightly smaller median distances found in coronal holes may indicate a difference in network size compared to quiet Sun regions. Observations and transition-region imaging show that magnetic network features are smaller / less distinct in coronal holes than in quiet Sun regions \citep[][]{1999hassler}. The horizontal size of convective flows increases with depth below the Sun’s surface. This means that small magnetic loops and internetwork elements observed near the {photosphere} are gradually transported and structured by increasingly larger convective motions deeper down. As a result, small loops tend to merge or accumulate along downdrafts, forming a hierarchical magnetic network, which may differ between coronal hole and quiet Sun regions \citep[][]{2009Nordlund}.\\

For coronal holes, we find that small loops are distributed throughout the entire coronal hole, except in regions of strong magnetic field concentrations (magnetic elements){, as exemplarily shown in Figures~\ref{fig:overview}, \ref{fig:overview_app1}, \ref{fig:overview_app2}, and \ref{fig:overview_app3}.}This is consistent with observations showing that the majority ($>90\%$) of the coronal hole area is covered by a weak, nearly balanced background field, while the open magnetic flux is concentrated in strong unipolar magnetic elements of the coronal holes dominant polarity \citep[see][]{Hofmeister2019, Heinemann2018b, 2019heinemann_catch}. These low-lying and particularly flat loops ($a \approx 0.1$) are found to cover this background field {(see Fig.~\ref{fig:lowloops} and Tab.~\ref{tab:combined_stats})}.

A distinct difference between the low loops present in coronal holes and those in quiet Sun regions is their dependence on the mean magnetic flux density of the respective region under study. With increasing magnetic flux density, the loop height decreases ($cc = 0.81$), whereas in quiet Sun regions the median loop height remains nearly constant independent of the strength of the field {(Fig~\ref{fig:metric_maxheight_low}b,e)}. We further found that this relationship also holds for all coronal holes that do not exhibit any modeled open field signature ($J=0$) in this study. \citet{2019heinemann_catch} found that the coverage of magnetic elements in coronal holes increases linearly with their mean magnetic flux density. It is further known that the open magnetic fields rooted in these elements, which are typically located at the edges and nodes of the magnetic network, expand rapidly to approximately fill the corona in a uniform manner \citep[][]{ cranmer2009, Wedemeyer-Boehm2009,Temmer2023,2025Alzate}. \citet{Heinemann2021_farside} found that the occurrence of chromospheric bright points, that is, loops formed above small bipolar structures, in coronal holes can be used to approximate their mean magnetic flux density. Combining the theoretical understanding of the vertical magnetic field structure of coronal holes \citep[see Fig.~16 in][]{Wedemeyer-Boehm2009} with the results of this study suggests that a stronger unipolar magnetic field (i.e., higher flux density) exerts greater magnetic pressure, causing the field to expand more rapidly at lower altitudes and compressing the low loops. This may explain why stronger coronal holes tend to appear darker in EUV emission, with \citet{2019heinemann_catch} finding a moderate correlation ($cc = 0.55$) between the mean intensity of coronal holes in $193$\AA\ and $|$B$_{\mathrm{signed}}|$. In the context of potential field modeling, this suggests that the source surface height, defined as the altitude at which field lines can be regarded as radial, should depend, among other factors, on the magnetic field density of the coronal hole. \\

\cite{2004wiegelmann} compared loop statistics for 12 coronal holes and 8 quiet Sun regions, finding that small loops in coronal holes are shorter and lower than those in quiet Sun regions, yet similarly abundant. This is consistent with our results. In their study, they also found that this trend extends to higher loops, although such loops occur in coronal holes at a much lower rate, opposite to our findings. It is worth noting that \cite{2004wiegelmann} used a non-linear force-free model with a different spatial resolution, and in our analysis we cannot clearly distinguish between high loops in coronal holes and open field lines. Nevertheless, when considering our full sample of loops, the overall trends remain in good agreement with their results. Similarly, \cite{2023Matkovic} found that coronal bright points in coronal holes are associated with smaller and less bright loop structures compared to those in quiet Sun regions.

\section{Summary and Conclusions}\label{s:sum}

We investigated the magnetic topology of 702 well-observed, low-latitude coronal holes using a PFSS model and compared the derived properties with quiet Sun regions. Our main findings are:

\begin{enumerate}
    \item Low loops are abundantly present in the weak, balanced background field in coronal holes and are statistically smaller and narrower than in quiet Sun regions.
    
    \item In coronal holes, the median height of low-lying magnetic loops is strongly correlated with the mean magnetic flux density ($cc = 0.81$). This indicates that, near the solar surface, the magnetic field structure in coronal holes is largely determined by flux density, whereas in quiet Sun regions, no such relationship is evident.
    
    \item Coronal holes contain a significant number of loops that close much higher in the corona than in quiet Sun regions, although it cannot be determined whether these loops are truly closed.
    
    \item Differences between the modeled magnetic structures of coronal holes and quiet Sun regions are evident, even in areas where the PFSS model with the chosen source surface height does not show open fields.
\end{enumerate}

These findings suggest that the commonly adopted standard source surface height in PFSS models is unlikely to provide a universally accurate representation of coronal holes \citep[e.g., this has also been suggested by][]{Schulz1978, 2020Kruse, 2020Panasenco}. Instead, the distinct, field-strength-dependent magnetic structure observed here indicates that each coronal hole may require its own, individually adjusted source surface height to achieve better agreement with observations. 

\begin{acknowledgments}
 The SDO image data are available by courtesy of NASA and the respective science teams. Inspired by the International Space Science Institute (ISSI, Bern) Team on “Magnetic open flux and solar wind structuring in interplanetary space” (2019-2022). SGH acknowledges funding from the Research Council of Finland (Academy Fellowship) [370747; RIB-Wind]. JP acknowledges funding from the Research Council of Finland (projects 343581, 364852 and 370793). This research was funded in whole, or in part, by the Austrian Science Fund (FWF) [10.55776/J4560]. For the purpose of open access, the author has applied a CC BY public copyright licence to any Author Accepted Manuscript version arising from this submission.
 SGH thanks Eleanna Asvestari and Stefan J. Hofmeister for productive discussion.
\end{acknowledgments}

\begin{contribution}

SGH conducted the analysis and led the manuscript preparation. JP provided the magnetic field model, and MT contributed valuable input. All authors participated in revising and editing the manuscript.


\end{contribution}

%



\appendix

\section{Additional Figures}

\begin{figure*}[h]
   \centering
   \includegraphics[width=1\linewidth]{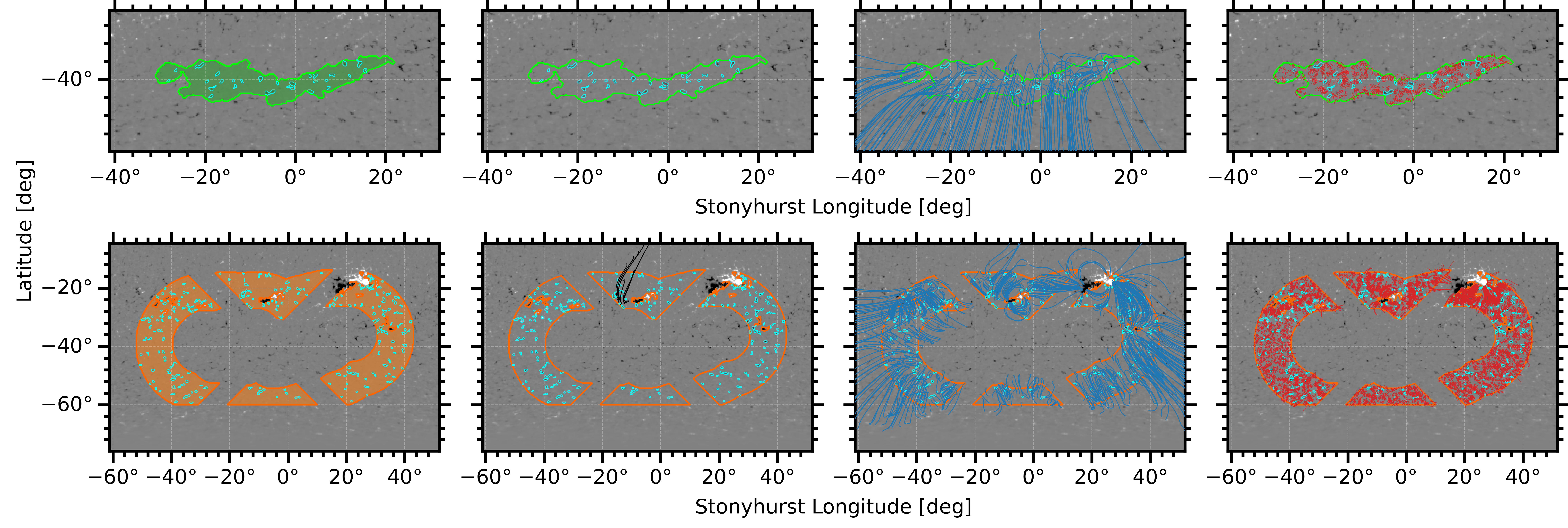}
   \caption{Same as Figure~\ref{fig:overview}, but for a coronal hole on February 13, 2012.}
              \label{fig:overview_app1}%
\end{figure*}

\begin{figure*}[h]
   \centering
   \includegraphics[width=1\linewidth]{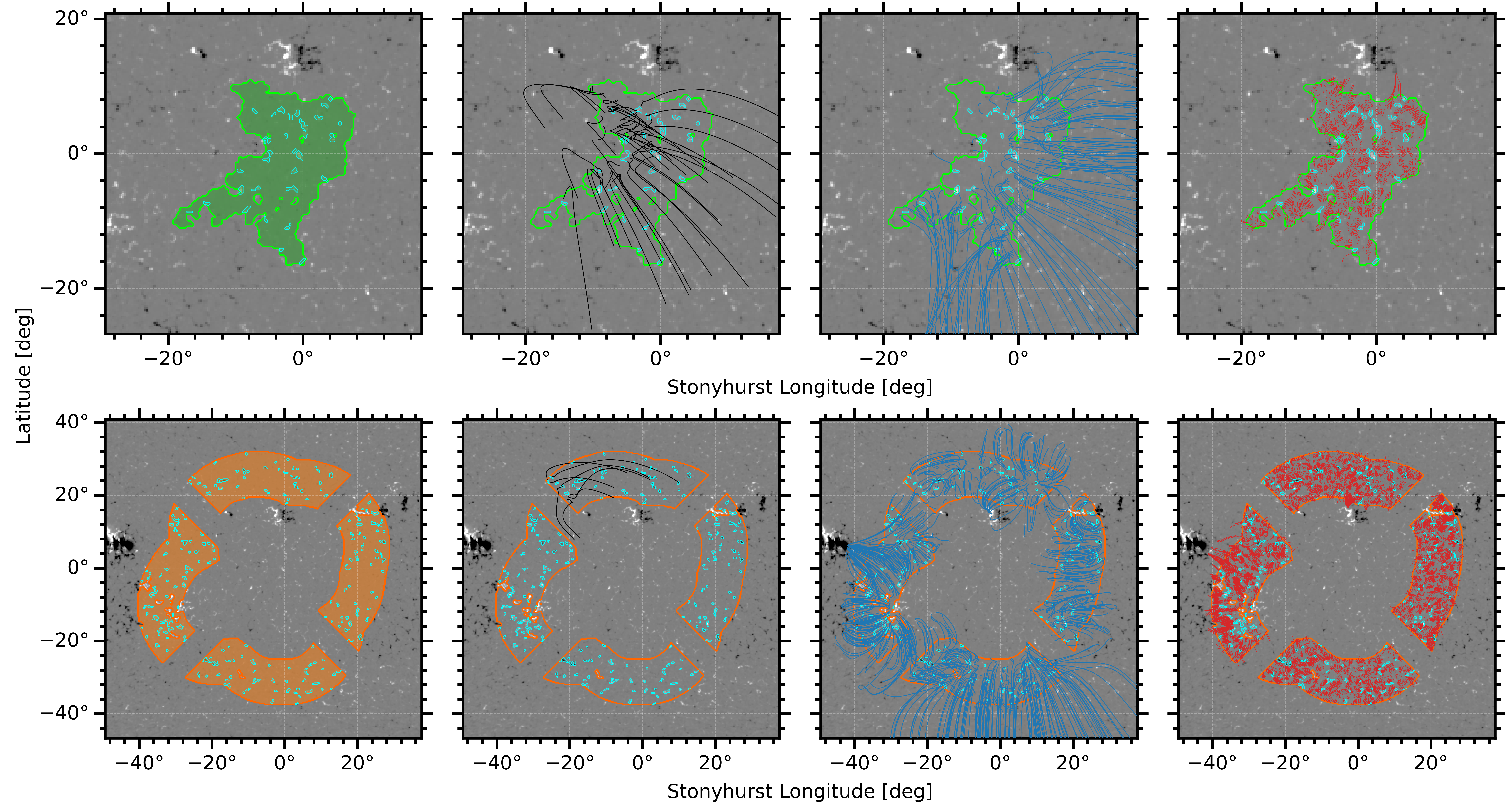}
   \caption{Same as Figure~\ref{fig:overview}, but for a coronal hole on October 31, 2015.}
              \label{fig:overview_app2}%
\end{figure*}

\begin{figure*}[h]
   \centering
   \includegraphics[width=1\linewidth]{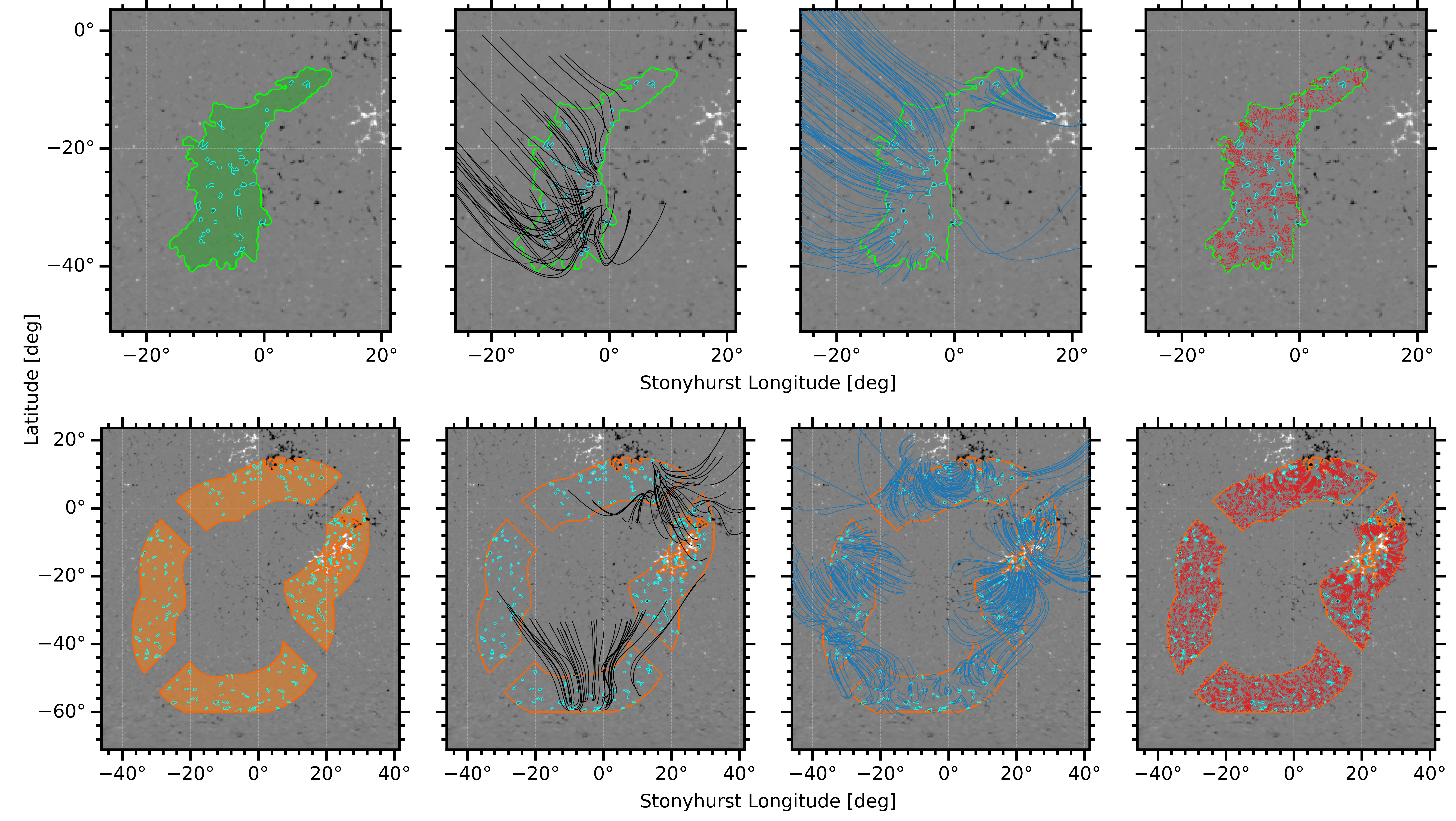}
   \caption{Same as Figure~\ref{fig:overview}, but for a coronal hole on October 29, 2017.}
              \label{fig:overview_app3}%
\end{figure*}

{In this appendix, we present additional figures. Figures~\ref{fig:overview_app1}, \ref{fig:overview_app2}, and \ref{fig:overview_app3} show three coronal holes together with their respective quiet-Sun regions used in the analysis of this study, analogous to Figure~\ref{fig:overview}. The extracted areas are overlaid on the magnetic maps used in the analysis, with open field lines shown in black and closed loops classified as high (blue) and low (red).}

\begin{figure*}[h]
   \centering
   \includegraphics[width=1\linewidth]{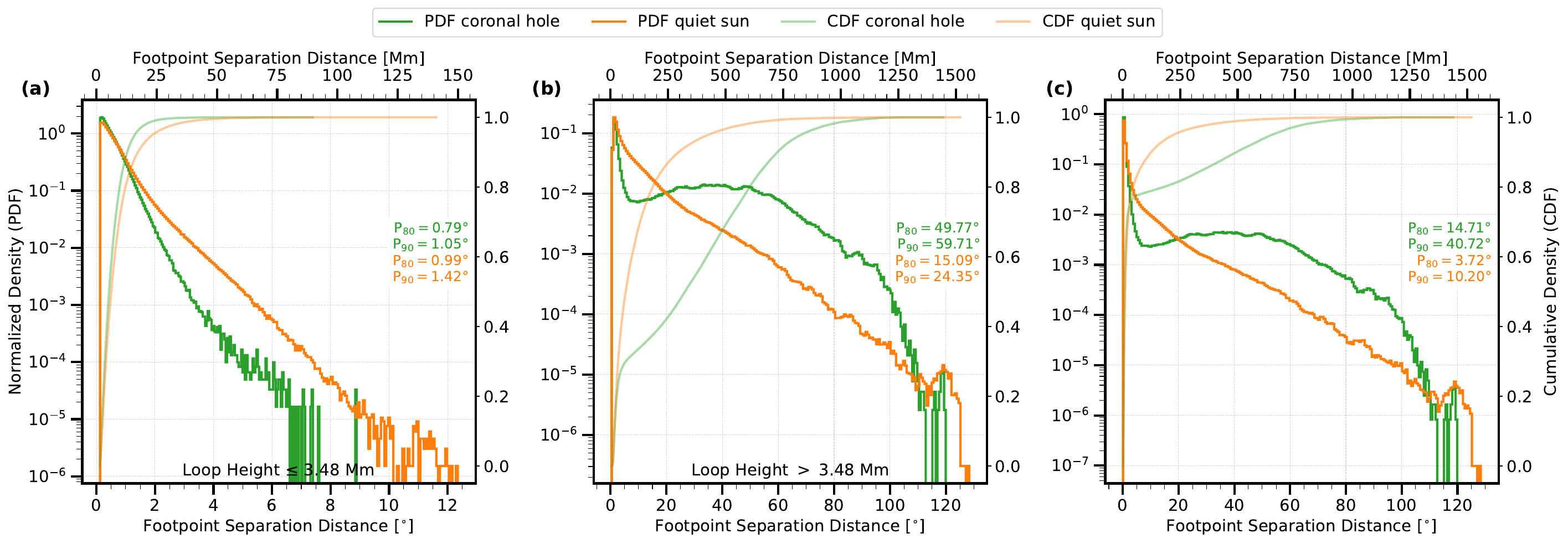}
   \caption{Same as Figure~\ref{fig:hist_maxheight} but for the loop footpoint separation distance.}
              \label{fig:hist_footpointsep}%
\end{figure*}

\begin{figure*}[h]
   \centering
   \includegraphics[width=1\linewidth]{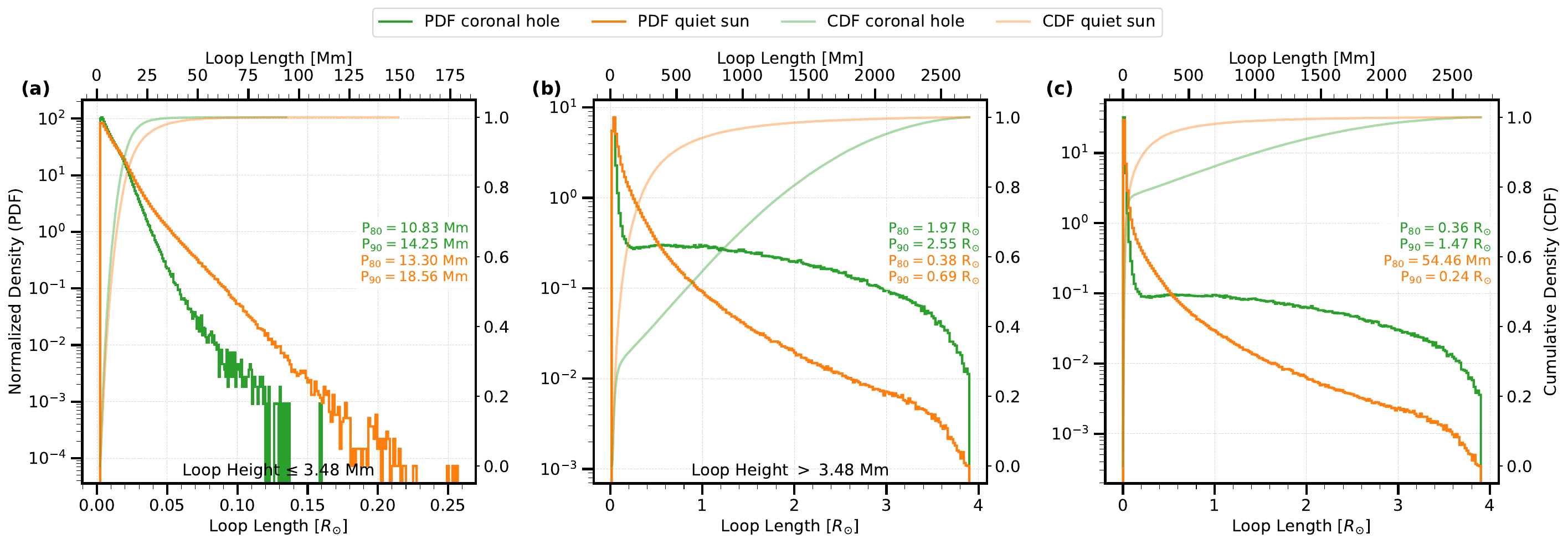}
   \caption{Same as Figure~\ref{fig:hist_maxheight} but for loop length.}
              \label{fig:hist_looplength}%
\end{figure*}

\begin{figure*}[h]
   \centering
   \includegraphics[width=1\linewidth]{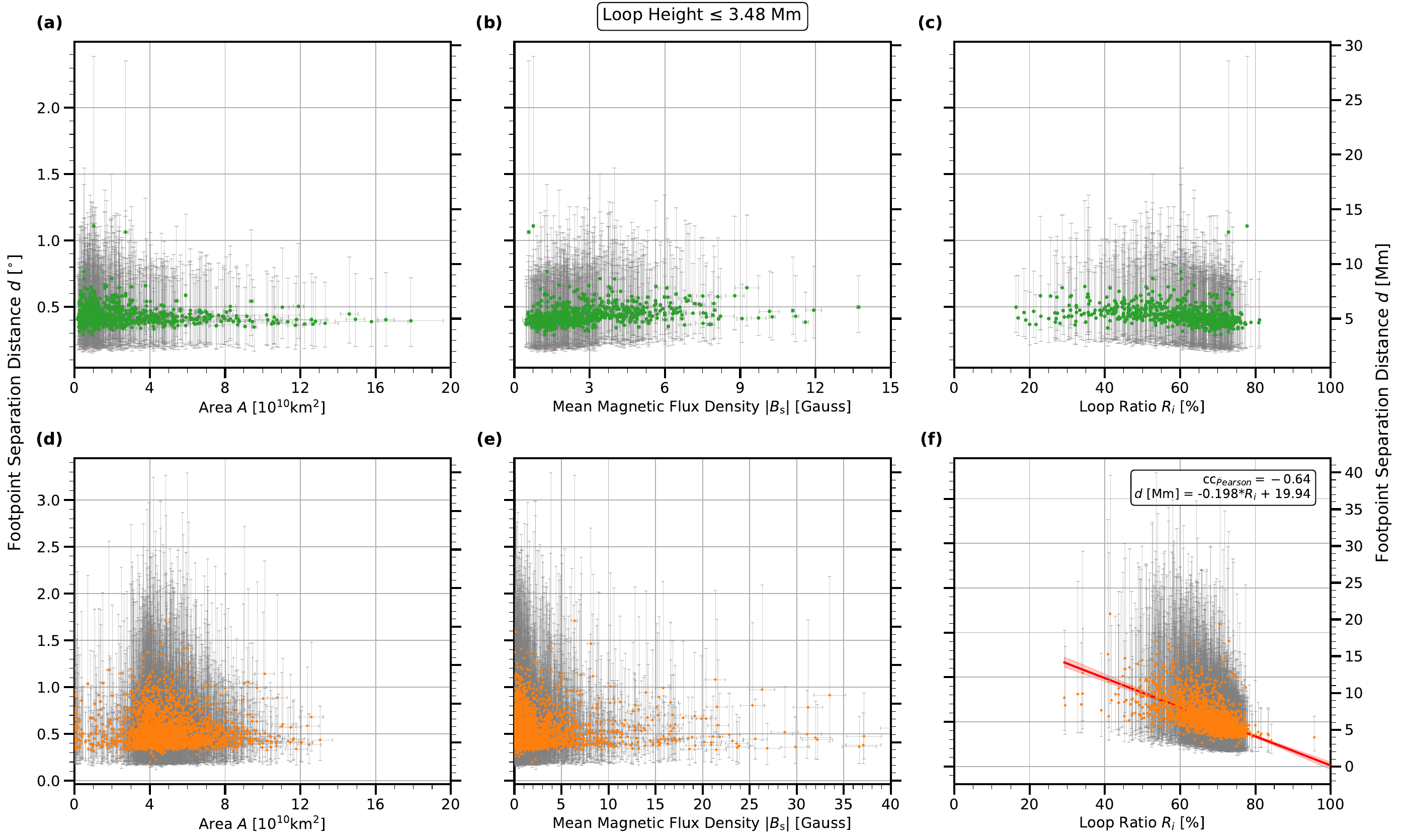}
   \caption{Same as Figure~\ref{fig:metric_maxheight_low} but for the loop footpoint separation distance.}
              \label{fig:metric_fp_distance_low}%
\end{figure*}

\begin{figure*}[h]
   \centering
   \includegraphics[width=1\linewidth]{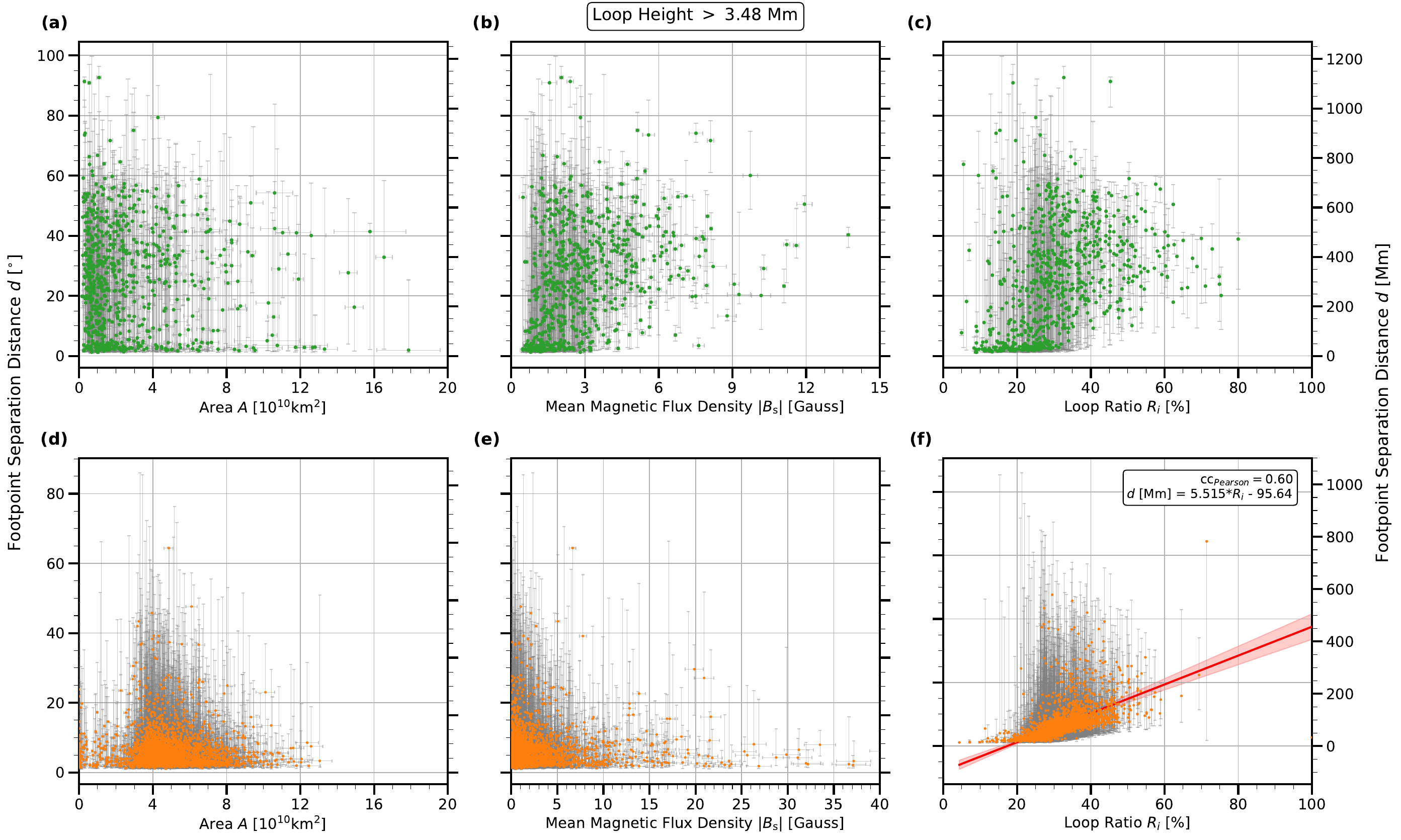}
   \caption{Same as Figure~\ref{fig:metric_maxheight_high} but for the loop footpoint separation distance.}
              \label{fig:metric_fp_distance_high}%
\end{figure*}

\begin{figure*}[h]
   \centering
   \includegraphics[width=1\linewidth]{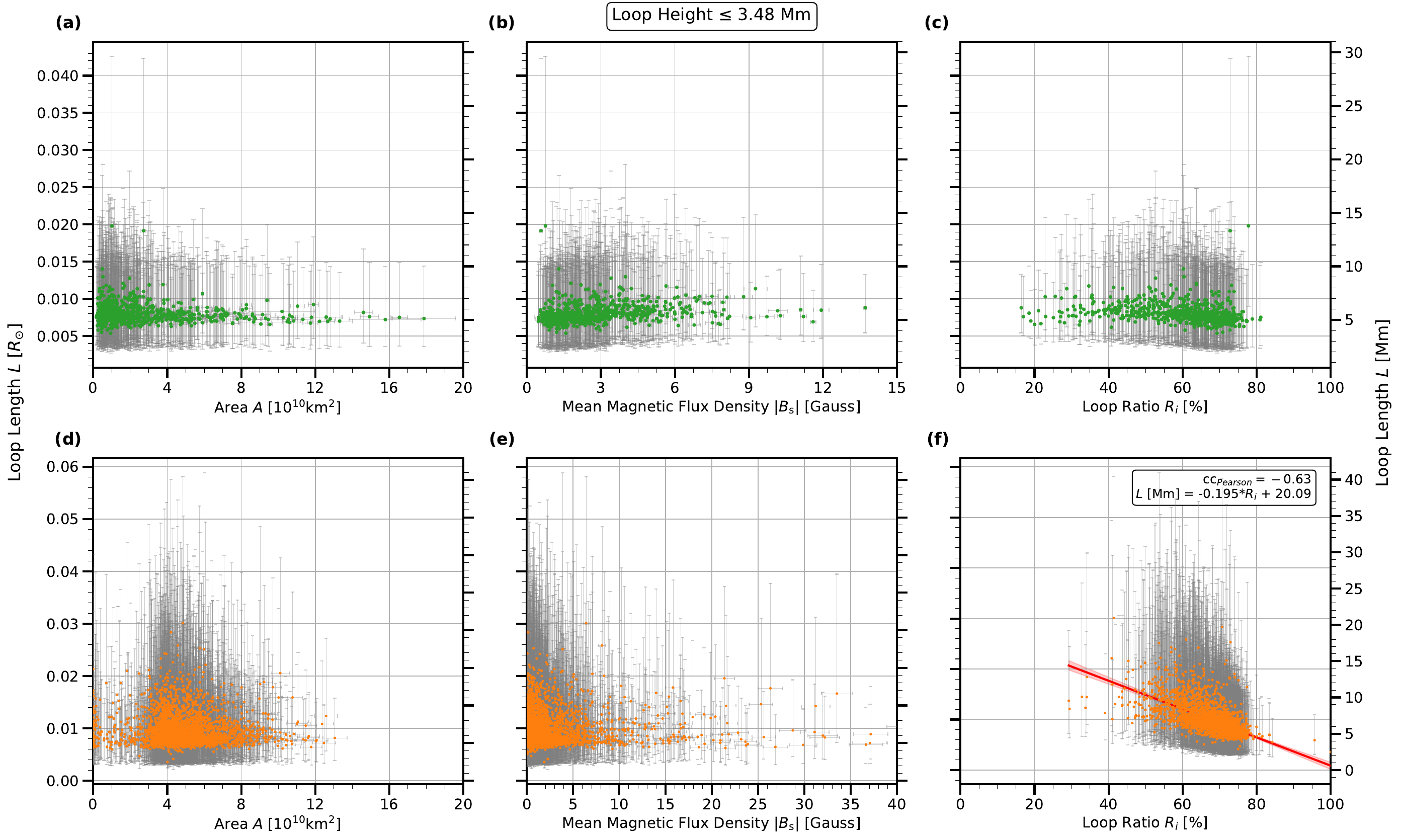}
   \caption{Same as Figure~\ref{fig:metric_maxheight_low} but for loop length.}
              \label{fig:metric_fl_length_low}%
\end{figure*}

\begin{figure*}[h]
   \centering
   \includegraphics[width=1\linewidth]{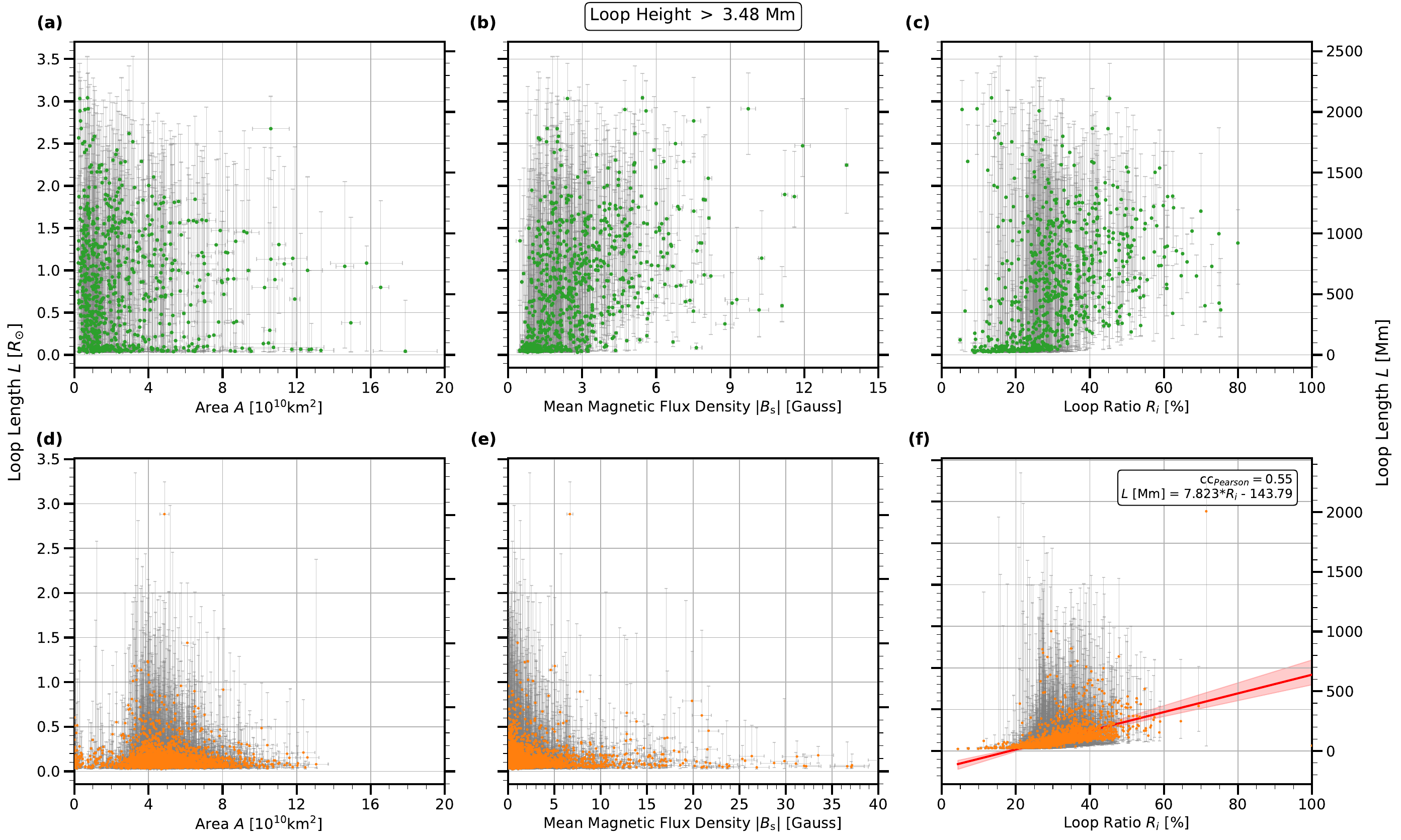}
   \caption{Same as Figure~\ref{fig:metric_maxheight_high} but for loop length.}
              \label{fig:metric_fl_length_high}%
\end{figure*}

{Figures~\ref{fig:hist_footpointsep} and~\ref{fig:hist_looplength} present histograms of the footpoint separation distance ($d$) and loop length ($L$) for low and high loops, including the 80th and 90th percentiles. Furthermore, Figures~\ref{fig:metric_fp_distance_low}, \ref{fig:metric_fp_distance_high}, \ref{fig:metric_fl_length_low}, and~\ref{fig:metric_fl_length_high} show scatter plots of $d$ and $L$ as functions of coronal hole and quiet-Sun region properties, namely the area ($A$), the absolute value of the mean magnetic flux density ($|B_s|$), and the loop ratio ($R_i$).}

\begin{figure*}[h]
   \centering
   \includegraphics[width=1\linewidth]{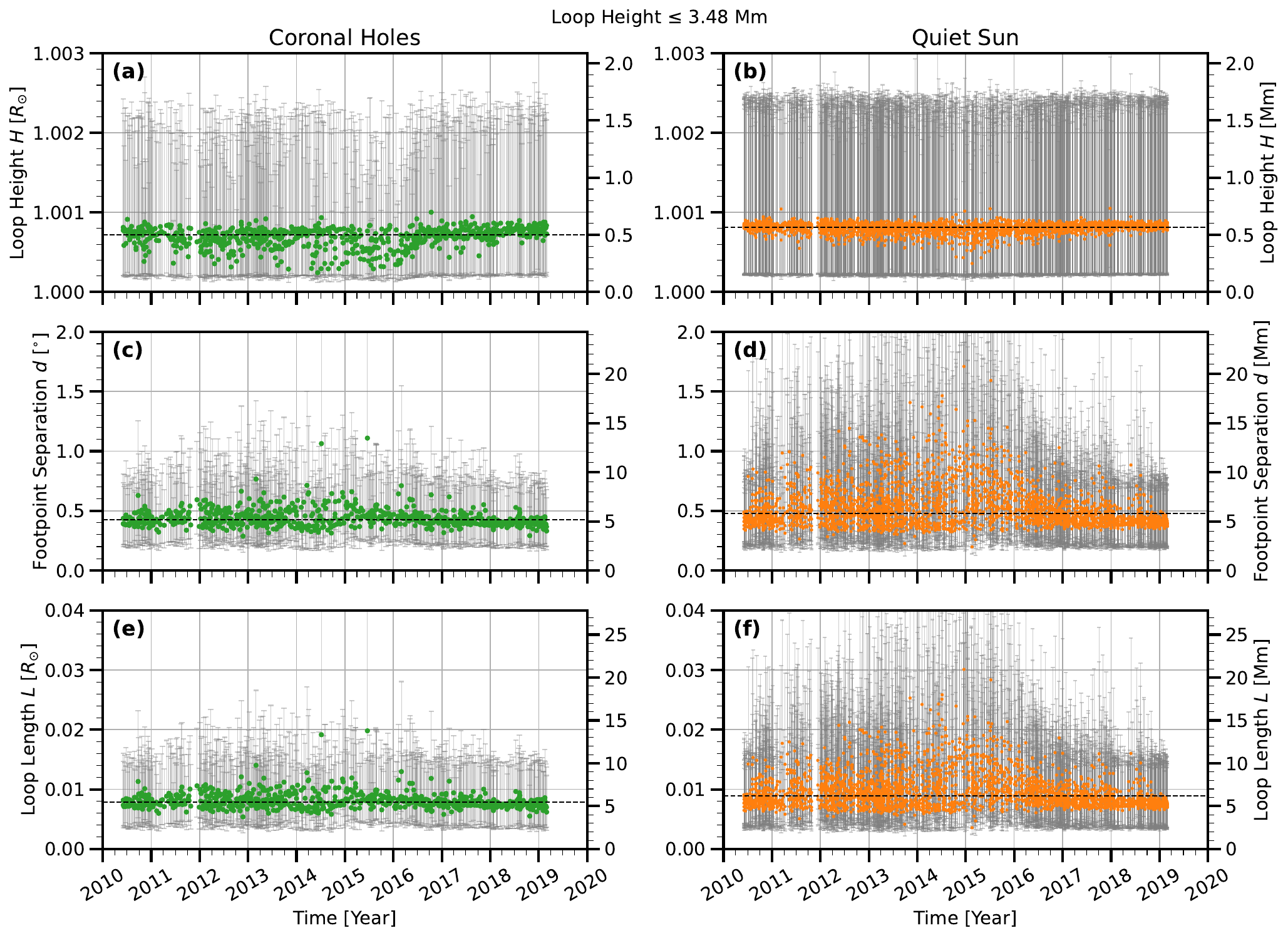}
   \caption{Median loop parameters ($H$, $d$, and $L$) of low loops ($H \le 3.48$~Mm) as a function of time for coronal-hole (a, c, e; green) and quiet Sun regions (b, d, f; orange) between 2010 and 2019. The grey shaded areas indicate the 20\textsuperscript{th} and 80\textsuperscript{th} percentiles for each region, and the dashed black line marks the overall median.}
              \label{fig:dt_low}%
\end{figure*}

\begin{figure*}[h]
   \centering
   \includegraphics[width=1\linewidth]{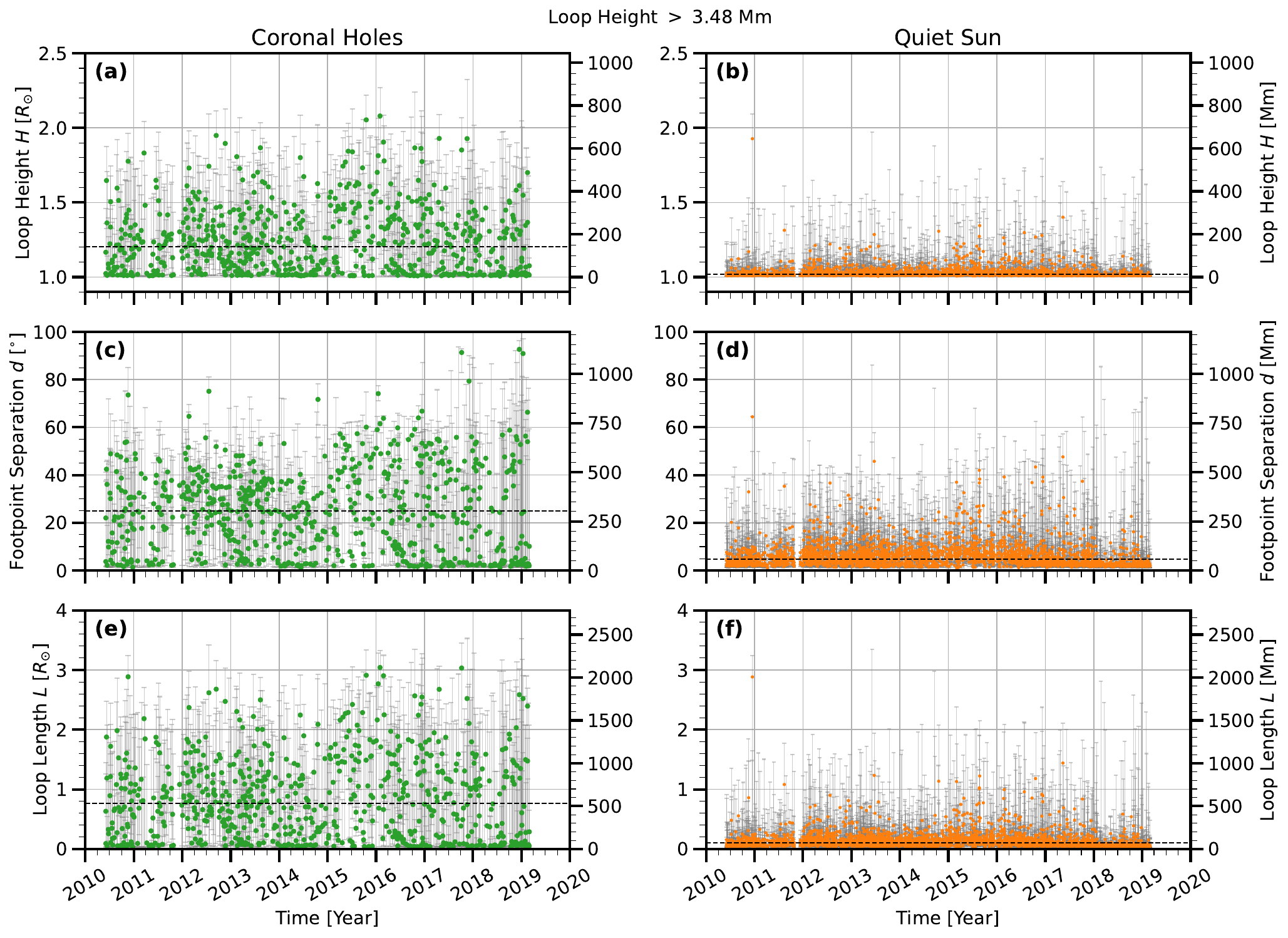}
   \caption{Same as Figure~\ref{fig:dt_low} but for high loops.}
              \label{fig:dt_high}%
\end{figure*}

\begin{figure*}[h]
   \centering
   \includegraphics[width=1\linewidth]{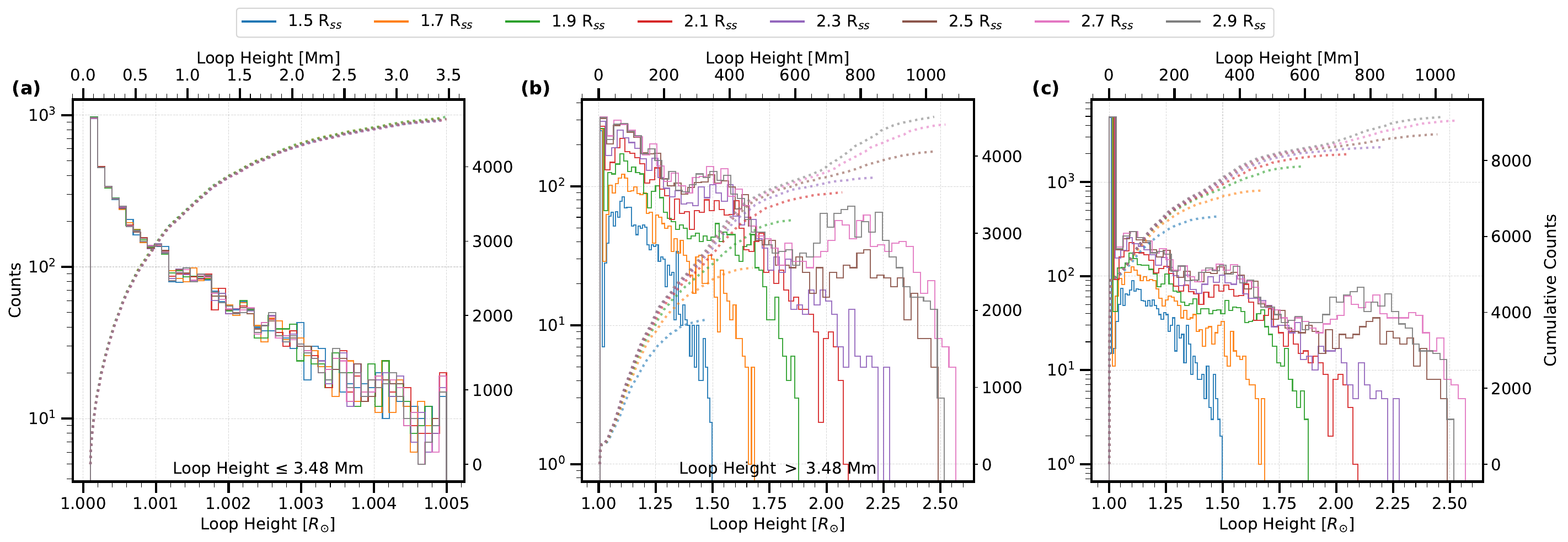}
   \caption{Distribution of the maximum loop height for low (a), high (b) and all loops (c) for the coronal hole on May 29, 2013, shown in Figure~\ref{fig:overview}, for different source surface heights from $1.5$ to $2.9$~$R_{\odot}$.  The solid line show the counts as function of loop height $H$ and the dotted lines the cumulative counts.}
              \label{fig:Rss_example}%
\end{figure*}

{In Figures~\ref{fig:dt_low} and~\ref{fig:dt_high}, we present the temporal evolution of the median values of the parameters $H$, $d$, and $L$ for each respective region. Finally, Figure~\ref{fig:Rss_example} shows the distribution of loop heights ($H$) for a single coronal hole observed on 29 May 2013 (shown in Figure~\ref{fig:overview}), for different source-surface heights.}

\bibliographystyle{aasjournalv7}



\end{document}